\documentclass[twocolumn,prl,preprintnumbers,amsmath,amssymb]{revtex4}
\usepackage{graphicx}% Include figure files
\usepackage{dcolumn}% Align table columns on decimal point
\usepackage{bm}% bold math
\usepackage{tabu}

\begin{document}
\title{Simple and accurate model of fracture toughness of solids}

\author{Haiyang Niu$^1$}

\email[]{haiyang.niu@phys.chem.ethz.ch}

\author{Shiwei Niu$^2$}
\email[]{nshiwei1123@gmail.com}

\author{Artem R. Oganov$^{3,4,5}$}
\email[]{a.oganov@skoltech.ru}

\affiliation{$^1$ Department of Geosciences and Center for Materials by Design, Institute for Advanced Computational Science, State University of New York, Stony Brook, NY, 11794-2100, USA.}
\affiliation{$^2$ Mining Technology Institute, Taiyuan University of Technology, Taiyuan, 030024, China}
\affiliation{$^3$ Skolkovo Institute of Science and Technology, Skolkovo Innovation Center, 143026, 4 Nobel Street, Moscow, Russia.} 
\affiliation{$^4$ Moscow Institute of Physics and Technology, 141700, 9 Institutsky lane, Dolgoprudny, Russia.}
\affiliation{$^5$ International Center for Materials Discovery, Northwestern Polytechnical University, Xi’an, 710072, China.}

\date{\today}
\begin{abstract}

Fracture toughness $K_{IC}$ plays an important role in materials design. Along with numerous experimental methods to measure fracture toughness of materials, its understanding and theoretical prediction is very important. However, theoretical prediction of fracture toughness is challenging. By investigating the correlation between fracture toughness and elastic properties of materials, we have constructed a fracture toughness model for covalent and ionic crystals. Furthermore, by introducing an enhancement factor, which is determined by the density of states at the Fermi level and atomic electronegativities, we have constructed a universal model of fracture toughness for covalent and ionic crystals, metals and intermetallics. The predicted fracture toughnesses are in good agreement with experimental values for a series of materials. All the parameters in the proposed model of fracture toughness can be obtained from first-principles calculations, which makes it suitable for practical applications. 
\end{abstract}

%\pacs{62.20.-x, 71.20-b, 71.15.Mb}

\maketitle

Fracture toughness $K_{IC}$ measures the resistance of a material against crack propagation, and is one of the most important mechanical properties of materials~\cite{ashby2004}. For example, materials used in drilling bits or in ballistic vest should posses not only high hardness, but also high fracture toughness. The most widely used materials, diamond and tungsten carbide (WC), have drawbacks. Diamond is expensive and has problems with chemical and thermal stability, while WC is very dense (ruling out some applications) and is not superhard. Numerous methods have been employed to experimentally measure this property. Theoretical understanding and prediction of fracture toughness of materials have attracted enormous attention~\cite{ding2004,king2007,ogata2009}. A simple approach is to seek a correlation between cohesive energy and fracture toughness~\cite{king2007}. However, as we know, metals have much higher $K_{IC}$, while having lower cohesive energies than covalent and ionic crystals. Apparently, $K_{IC}$ is not a function of just the scalar cohesive energy. By introducing more detailed mechanical and quantum-mechanical attributes of bonding~\cite{ding2004,ogata2009}, such as the ideal strength, band gap, ionicity, etc., the correlation between bonding properties and fracture toughness of materials is improved but still insufficient for actual applications. 

Similar to fracture toughness, measuring hardness involves fracture and deformation under mixed loading conditions. There have been several attempts to establish correlations between hardness and one single elastic property such as bulk modulus or shear modulus~\cite{liu1989,teter1998,niu2012}. By combining shear modulus and the Pugh modulus ratio $B/G$~\cite{pugh1954}($B$ and $G$ refer to bulk and shear modulus, respectively), an empirical model for predicting hardness of polycrystalline material has been proposed by Chen et al.~\cite{chen2011}, and proved to be very reliable by many applications~\cite{zhang2013,lyakhov2011, zhang2014-1,szwacki2017}.
Furthermore, by introducing the concept of bond strength, other researchers have proposed robust hardness models for covalent~\cite{gao2003} and ionic crystals~\cite{simunek2006}.  
This inspired us to consider whether it is possible to construct a model of fracture toughness based on the elastic and electronic properties of materials.

In this Letter we propose a simple and accurate model of fracture toughness for covalent and ionic crystals using mainly elastic properties of materials. Considering the fracture toughness of metals is usually 1-2 orders higher than that of covalent and ionic crystals, we introduce an enhancement factor by combining the density of states at the Fermi level and the electronegativity of materials.  With that, we obtain a universal, simple and physically transparent model, working across three orders of magnitude and applicable to covalent and ionic crystals, metals and intermetallics. All the parameters in the model can be easily obtained by first-principles calculations, which makes the model applicable to materials selection and design.

Theoretical stress intensity factor to propagate a crack for materials with a crack under mode I loading (the load is normal to the cleavage plane) is given by Griffith theory~\cite{griffith1921}. Through breaking atomic bonds a crack propagates, and consequently new surfaces are generated. The surface tension of the opening surfaces at the crack tip 2$\gamma_s$ ( $\gamma_s$ is the surface energy of the material) is the force to balance this elastic driving force. Therefore, the critical value of the stress intensity factor under mode I loading is given by:
\begin{equation}\label{E:stressintensity}
  K_g=2\sqrt{\gamma_sG/(1-\nu)} \,,
\end{equation}
where $\nu$ is Poisson's ratio, and $G$ is shear modulus. This equation is often called the Griffith relation~\cite{griffith1921}. When the stress intensity factor of a crack reaches $K_g$, a crack propagates. $K_g$ is the so-called theoretical fracture toughness~\cite{thomson1987}, which is applicable to materials without defects. In practice, an experimentally measured fracture toughness $K_{IC}$ is considerably lower than $K_g$, making Eq.~(\ref{E:stressintensity}) not useful in practice. However, the underlying physical correlation between fracture toughness and elastic properties provides useful insight. 

\begin{table*}
 \centering
 \caption{Comparison between predicted fracture toughness $K_{IC}$ with available experimental values at room temperature for a series of covalent and ionic crystals, along with predictions for some materials, such as CrB$_4$, $\gamma$-B$_{28}$ and Fe$_3$C. The calculated shear modulus $G$, bulk modulus $B$, volume per atom $V_0$, and Pugh modulus ratio $B/G$ are also given. Experimental $K_{IC}$ obtained according to ASTM standards are taken from Ref.~\cite{nist1998} unless otherwise specified.  Experimental $H_v^{exp}$ are taken from Ref.~\cite{chen2011} and Ref.~\cite{gao2003} unless otherwise specified. Predicted $H_v^{pre}$ are estimated by Chen's model~\cite{chen2011}. Elastic properties and volume per atom $V_0$ of materials are calculated within the framework of density functional using the PBE exchange-correlation functional~\cite{perdew1996} within the generalized gradient approximation and the projector-augmented waves method~\cite{blochl1994} as implemented in VASP~\cite{kresse1993,kresse1996}. The calculated bulk ($B$) and shear ($G$) moduli are determined with Reuss-Voigt-Hill approximation~\cite{hill1952}}.
 \label{T:parameters}
 \begin{tabular} {cccccccccc}
 \hline\hline
 Material & $G$ & $B$ & $V_0$ & $B/G$ & $K_{IC}^{exp}$ & $K_{IC}^{pre}$  & $H_v^{exp}$ & $H_v^{pre}$\\
  & GPa & GPa & $\mathring{A}$$^3$/atom & & MPa$\cdotp$$m^{1/2}$ & MPa$\cdotp$$m^{1/2}$ & GPa & GPa \\
 \hline
 diamond    & 520.3 & 431.9 & 5.70 & 0.83 & 5.3, 6.6, 6.7     & 6.33  & 96 & 93.5\\
 \hline
    WC  & 301.8 & 438.9 & 10.61  &1.46& 7.5~\cite{nist1998} & 5.40 & 24, 30 &29.3 \\
  \hline
 BN         & 403.4 & 403.7 & 5.95  & 1.00 & 5               & 5.97 & 66& 63.8\\
 \hline
   TiN & 183.2   & 282   & 9.66  &1.54 & 3.4, 4.28, 5.0~\cite{nist1998}& 3.32 & 23&22.5 \\
  \hline
  WB$_3$& 220.1& 307.2& 8.79& 1.40 & & 3.73& 28.1-43.3~\cite{cheng2014} & 28.8\\
  \hline
  CrB$_4$& 261.0& 265.3& 7.45& 1.01 & & 3.68& &48\\
  \hline
  SiO$_2$$^a$ & 220.0  & 305.0   & 7.75 & 1.37 &                & 3.64 & 33&30.4 \\
 \hline 
   TiC & 176.9 & 250.3 & 10.19 &1.42& 2-3,3.8~\cite{nist1998} & 3.10 & 24.7&24.5 \\
  \hline
  $\gamma$-B$_{28}$ & 236.0 & 224.0 & 6.99  & 0.95 &             & 3.18 & 50& 49.0\\
 \hline
 B$_6$O & 204.0    &228.0   & 7.39 & 1.12          &             & 3.01 & 38 & 36.4 \\
 \hline
 SiC        & 196.6 & 224.9 & 10.49 & 1.14 &3.1, 3.3, 4.0            & 3.11 & 34&34.5 \\
 \hline
  Al$_2$O$_3$& 164.3   & 254.2   & 8.75 & 1.55 & 3~\cite{weber2002}               & 2.93 & 20&20.6 \\
 \hline
 B$_4$C     & 191.9  & 225.8   & 7.42  & 1.18 & 3.08, 3.2, 3.7         & 2.91 & 30&32.8\\
 \hline
    AlN        & 122.1   & 194.1   & 10.63 & 1.59 & 2.79            & 2.28 & 18 & 16.3\\    
  \hline
   TiO$_2$ & 110.1& 209.2   & 12.22 & 1.90 & 2.1~\cite{anstis1981}, 2.8         & 2.30 & &11.7 \\  
  \hline
  $\alpha$-Si$_3$N$_4$& 120.1&233.8&10.62 & 1.95 & 3.12~\cite{tanaka1994}         & 2.48 & &12.1\\  
  \hline
 MgO        & 130.3 & 158.3 & 9.67  & 1.21 & 1.9, 2.0         & 2.09 & &24.5 \\
   \hline
 ThO$_2$     & 88.1    & 187.7   & 14.79 & 2.14 & 1.07            & 2.01 & &8.3\\  
  \hline
   Fe$_3$C     & 81.5   & 223.2   & 9.51  & 2.74 &          & 1.96 & &5.1\\
 \hline
MgAl$_2$O$_4$& 96.1   & 180.2   & 9.73  & 1.88 & 1.83, 1.94, 1.97           & 1.92 & &10.8\\  
  \hline
  Y$_2$O$_3$  & 61.3    & 138.5   & 15.33 & 2.26 & 0.71            & 1.45 & 7.5&5.5\\  
   \hline 
ZnO$_2$     & 62.1    & 113.8   & 10.15 & 2.32 & 1.6, 2.5         & 1.39 & &5.3\\  
  \hline 
 Si         & 66.3  & 98.2    & 20.41 & 1.48 & 0.79, 0.95  & 1.33 &12 & 11.7 \\
  \hline
 GaP        & 55.8  & 88.8  & 21.18 & 1.59 & 0.9~\cite{weber2002}            & 1.17 & 9.5 &9.2 \\
  \hline
 Ge         & 53.1  & 72.2  & 24.17 & 1.36 & 0.59-0.64~\cite{lemaitre1988} & 1.05 & 11.2 &8.8 \\
 \hline
 MgF$_2$     & 52.2    & 95.3    & 11.36 & 1.83 & 0.98            & 1.05 & &6.9\\  
  \hline
 GaAs       & 46.7  & 75.5  & 23.92 & 1.62 &0.44~\cite{michot1988} &1.01 & 7.5&7.8\\
 \hline
  BaTiO$_3$  & 45.1    & 94.9    & 13.15 & 2.11 & 1.05            & 1.01 &  & 4.7 \\  
  \hline
  InP        & 34.3  & 72.5  & 26.99 & 2.11 & 0.42-0.53~\cite{ericson1988}       & 0.86 & 5.4&3.6\\
 \hline
 ZnS        & 32.8  & 78.4  & 20.21  & 2.39 & 0.75, 1.0               & 0.84 & 1.8& 2.5\\
 \hline 
 ZnSe        & 28.1    & 58.4    & 23.60  & 2.07 & 0.32~\cite{weber2002}, 1~\cite{weber2002} & 0.68 & 1.4&2.9\\  
  \hline
  CdS        & 18.6 & 61.1  & 26.07 &3.28 & 0.33-0.76~\cite{tafreshi1995}     & 0.58 \\
  \hline     
 CdSe       & 16.3 & 53.1  & 29.79 &3.26 & 0.33-1.2~\cite{tafreshi1995}         & 0.52 \\
 \hline
 NaCl       & 14.8 & 24.9 & 22.61 & 1.69 & 0.17-0.22~\cite{narita1987} & 0.32 & 0.3 & 2.2\\
  \hline\hline
$^a$Stishovite.\\
\end{tabular}
\end{table*}

\begin{figure}
  \centering
  \includegraphics[width=0.85\columnwidth]{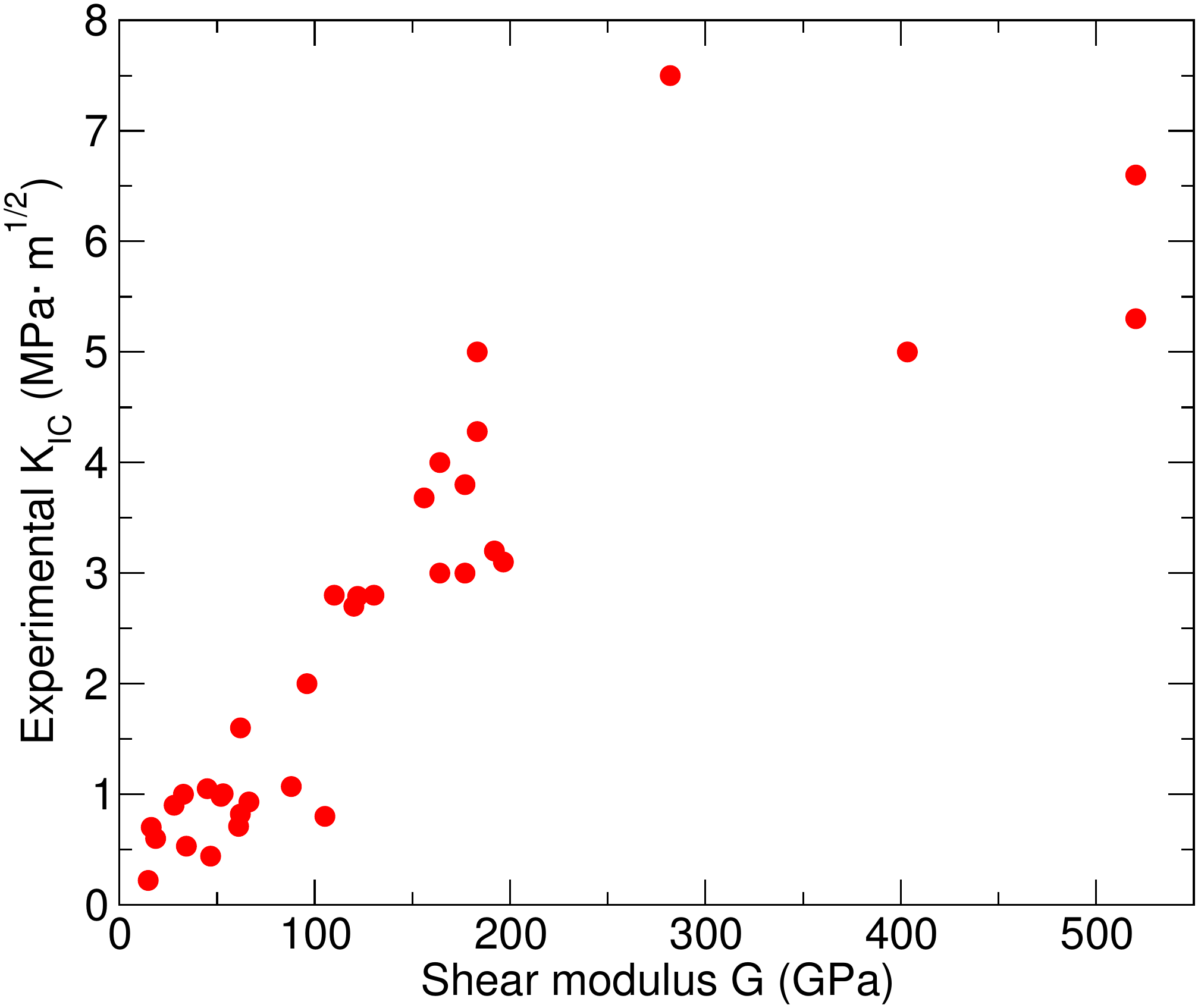}
  \caption{Correlation between shear modulus $G$ and experimental fracture toughness $K_{IC}$.}
  \label{F:f1}
\end{figure}

In order to evaluate the correlation between shear modulus $G$ and experimental fracture toughness $K_{IC}$, we plot in Fig. 1 experimental $K_{IC}$ against shear modulus $G$ for a series of covalent and ionic crystals. From Fig. 1 we can see that the correlation between $K_{IC}$ and $G$ is not linear, but in general $K_{IC}$ increases with $G$ with the correlation coefficient between them of 0.90. Therefore, it is promising to get a model of fracture toughness by adding some correction to the correlation between $K_{IC}$ and $G$.

One possible correction factor is the well-known Pugh modulus ratio $B/G$~\cite{pugh1954}. Pugh~\cite{pugh1954} found that $B/G$ is closely related with brittleness and ductility of materials. The lower the value of $B/G$, the more brittle a material would be. Importantly, Pugh~\cite{pugh1954} also highlighted that the critical strain at fracture can be measured as $\varepsilon \propto (B/G)^2$. During deformation of a material, bonds break and reform resulting in displacement of atoms and slipping of atomic planes, and materials with high fracture toughness usually exhibit high ductility and yield at high critical strain. Therefore, we conclude that $B/G$ is in positive correlation with fracture toughness $K_{IC}$.

\begin{figure}
  \centering
  \includegraphics[width=0.85\columnwidth]{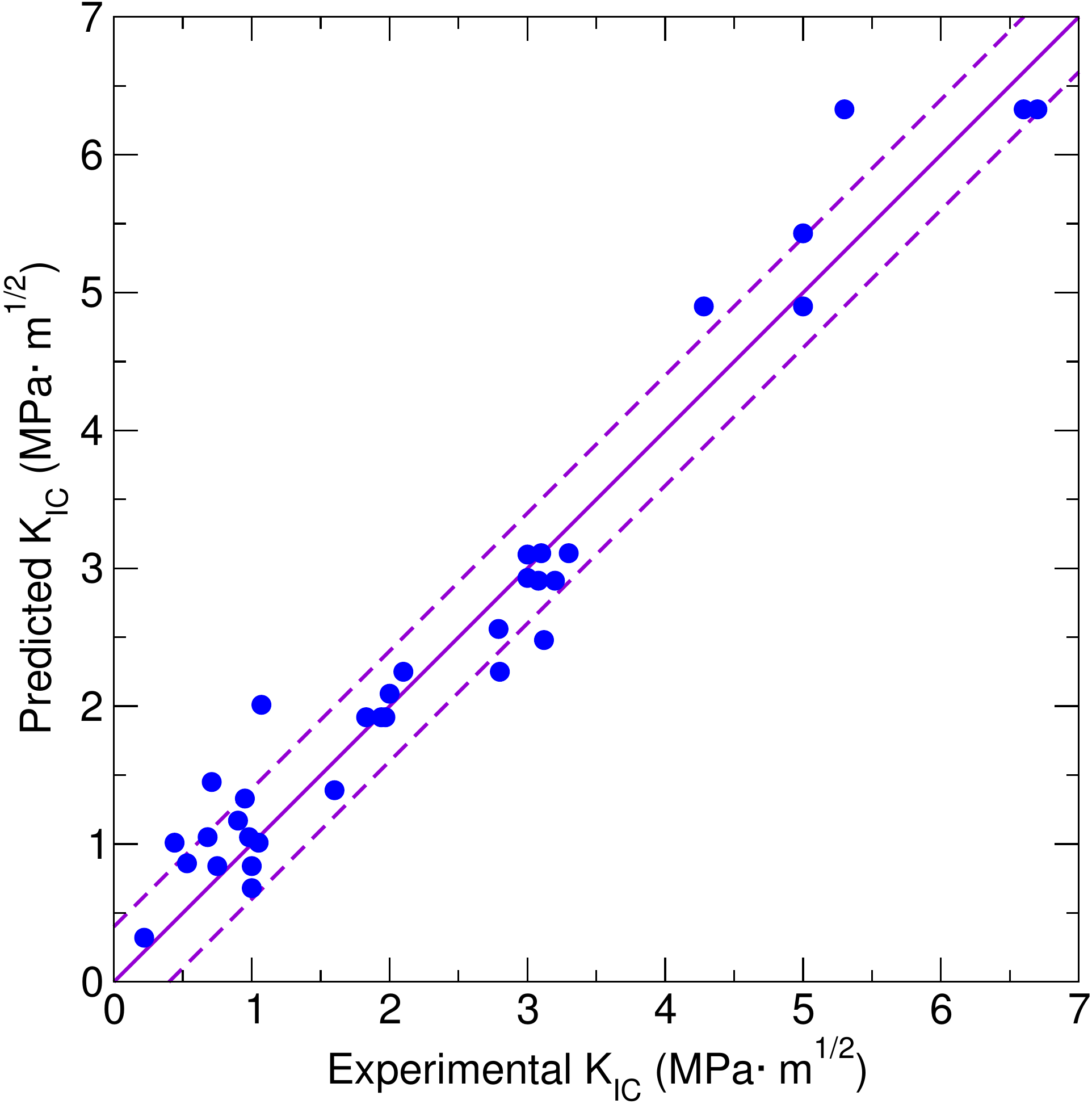}
  \caption{Comparison between experimental and predicted fracture toughness $K_{IC}$ for a series of ceramics. The root mean square error (RMSE) between experimental and predicted values are drawn with dotted line to guide the eyes.}
  \label{F:f2}
\end{figure}

By combining shear modulus $G$ and Pugh modulus ratio $B/G$, we propose the following empirical relation:
\begin{equation}\label{E:KIC1}
  K_{IC}\propto G \cdotp (B/G)^m \,.
\end{equation}

In order to have correct dimensionality of $K_{IC}$ (MPa $\cdotp$ $m^{1/2}$), a length scale unit must be added. Here, we have added volume per atom $V_0$ to the above relation, and by fitting to the data in Table I, we can get the value of m to be about 0.5. Thus we can obtain the following empirical formula for calculating fracture toughness of covalent and ionic crystals:
\begin{equation}\label{E:KICinsulator}
  K_{IC}=V_0^{1/6} \cdotp G \cdotp (B/G)^{1/2} \,.
\end{equation}
where $V_0$ is the volume per atom (in $m^3$), G and B are shear and bulk moduli (in MPa), and the unit of $K_{IC}$ is in MPa$\cdotp$$m^{1/2}$. The comparison between the predicted and experimental fracture toughness is graphically represented in Fig. 2. As we can see, the predicted values are in agreement with experimental results. By calculating the correlation coefficient between predicted and experimental value, we have found this value to be 0.97. Furthermore the root mean square error (EMSE) is estimated to be about 0.4 MPa $\cdotp$ $m^{1/2}$. The high correlation coefficient and small RMSE indicate the reliability and robustness of the proposed model of fracture toughness.

The fracture toughness of metals is usually 1-2 orders higher than that of ionic or covalent crystals, which is due to the lower crack sensitivity of metallic bonding compared with ionic and covalent bonds. Metallic bonds can be easily broken and reformed, while ionic and covalent bonds are very hard to break, but once broken, very hard to reform. If we use Eq.~(\ref{E:KICinsulator}) directly to calculate the fracture toughness of a metal, the resulting $K_{IC}$ are much lower than the experimental values. Considering the intrinsic difference between ceramics and metals, we can introduce an enhancement factor $\alpha$ to Eq.~(\ref{E:KICinsulator}) and obtain the following formula for metals:
\begin{equation}\label{E:KICmetal1}
  K_{IC}=(1+\alpha) \cdot V_0^{1/6} \cdotp G \cdotp (B/G)^{1/2} \,.
\end{equation}

The enhancement factor $\alpha$ shall distinguish between covalent and ionic crystals and metals and reflect the degree of metallicity. One choice could be the density of states (DOS) at the Fermi level. Importantly, we sum up the spin-up and spin-down electron DOS for magnetic materials to get the total DOS at the Fermi level. In order to correct the dimensionality of DOS, we need to choose a reference scale and calculate the relative DOS per volume at the Fermi level. Here we use free electron gas as the reference. With taking aluminum's atomic volume and valence electrons, we can get the DOS at the Fermi level of the free electron gas $g(E_F)_{FES}$ = 0.025 states$\cdotp$$eV^{-1}\mathring{A}^{-3}$. Thus the relative DOS at the Fermi level $g(E_F)_R$ = $g(E_F)$/$g(E_F)_{FES}$ of any metal can be obtained accordingly (See Table II).

 \begin{figure}
   \centering
   \includegraphics[width=0.8\columnwidth]{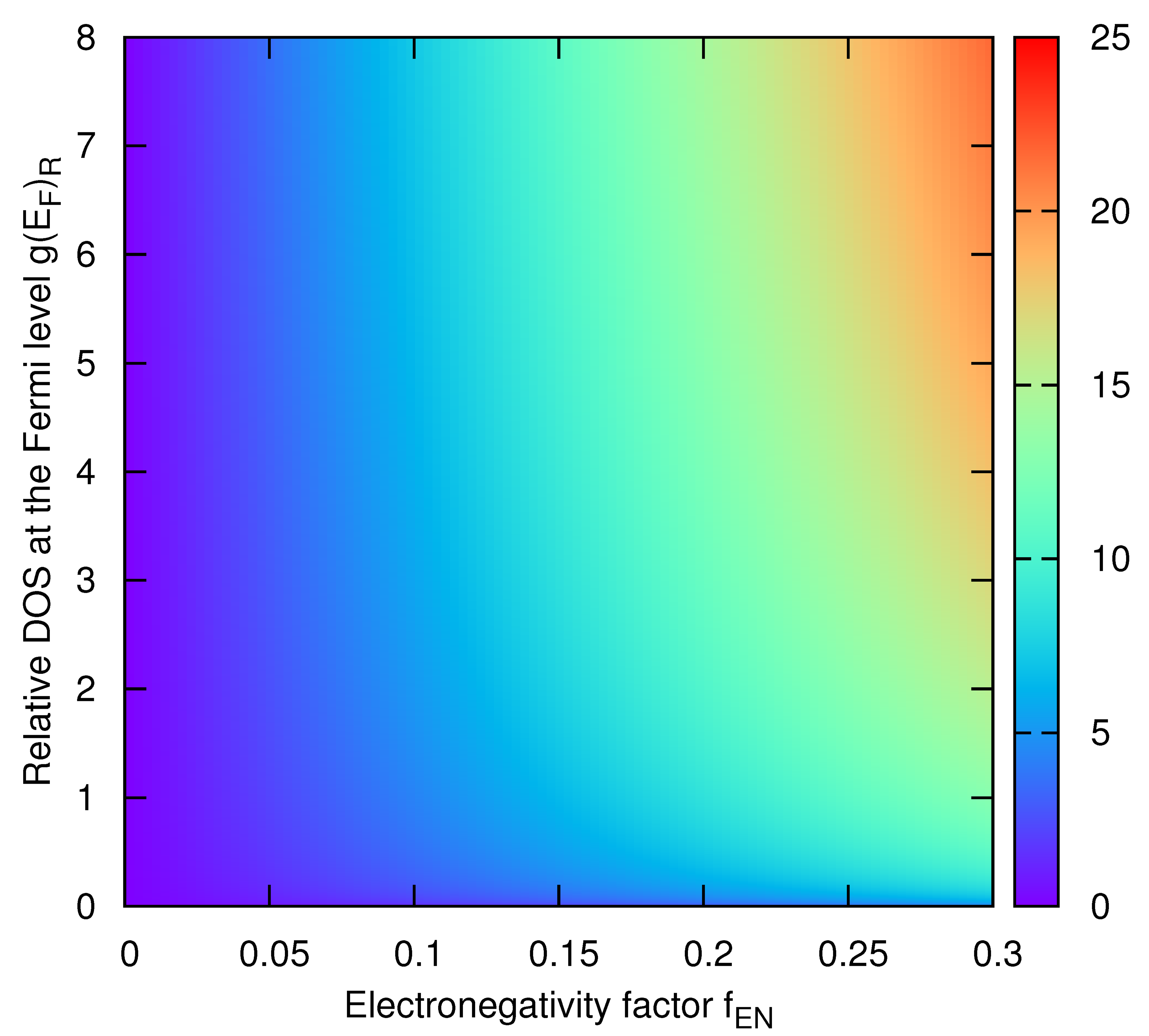}
   \caption{Correlation between the enhancement factor $\alpha$ and the electronegativity factor $f_{EN}$ and the relative DOS at the Fermi level $g(E_F)_R$. }
   \label{F:f3}
 \end{figure}

\begin{table*}
 \centering
 \caption{Comparison between predicted fracture toughness $K_{IC}$ with available experimental values at room temperature for a series of metals and intermetallics. The shear modulus $G$, bulk modulus $B$, volume per atom $V_0$, Pugh modulus ratio $B/G$, and the density of states at the Fermi level are also given. Allen scale electronegativity $\chi$~\cite{allen1989} of element A and B of compound A$_m$B$_n$ are also listed. $^a$ refers to ferromagnetic phase of Fe. $^b$ and $^c$ refer to Cu-Sn (3\%Sn) and Cu-Sn(9\%Sn) bronze, which are constructed by replacing 1 and 3 Cu atoms with Sn atoms in a 2$\times$ 2 $\times$ 2 supercell of Cu FCC lattice, respectively. $^d$~refers to tin bronze. Elastic properties, electronic properties and volume per atom $V_0$ of materials are calculated within the framework of density functional using the PBE exchange-correlation functional~\cite{perdew1996} within the generalized gradient approximation and the projector-augmented waves method~\cite{blochl1994} as implemented in VASP~\cite{kresse1993,kresse1996}. The calculated bulk ($B$) and shear ($G$) moduli are determined with Reuss-Voigt-Hill approximation~\cite{hill1952}. }
 \label{T:table2}
 \begin{tabular} {ccccccccccc}
 \hline\hline
 Material & $G$ & $B$ & $V_0$ & $\alpha$ & $B/G$ & g(E$_F$)$_R$ & $\chi_A$ & $\chi_B$ & $K_{IC}^{exp}$ & $K_{IC}^{pre}$  \\

     & GPa & GPa & $\mathring{A}$$^3$/atom & & & & & & MPa$\cdotp$$m^{1/2}$ & MPa$\cdotp$$m^{1/2}$ \\
 \hline
 Mg  & 24.3 & 44.6 & 22.87 & 35.3 & 1.84 & 0.45& - & - & 16-18 & 20.1 \\
   \hline
 Al  & 27.6 & 75   & 16.47 & 44.2 & 2.72 & 1.12 & - & - & 30-35 & 32.8  \\
   \hline
 V  & 37.1& 179.2& 13.41  & 65.9 & 5.52 & 4.84 & - & - & 70-150 & 84.0 \\
  \hline
 Ti  & 44.8 & 110.8& 17.05 &52.3 & 2.47 & 2.18 & - & - & 50-55 & 60.2\\
  \hline
 Ni  & 82.3 & 180.8& 10.78 &68.6 & 2.20 & 6.48 & - & - & 100-150& 126.2 \\
  \hline
 W   & 158.0  & 309.7& 15.91&43.3 & 1.94 & 1.02 & - & - & 120-150& 155.2 \\
  \hline
 Fe$^a$& 94.1& 193.9& 11.97 &59.2 & 2.06& 3.60 & - & - & 120-150& 123.1 \\
  \hline
 Ag  & 29.6  & 103.8 & 17.67&36.9 &3.51 &0.54 & - & - & 40-105 & 34.1 \\
  \hline
 Au  & 27.5  & 171.7 & 17.85 &37.2 &6.24&0.56 & - & - & 40-90  & 42.5 \\
  \hline
  $\beta$-Sn  & 21.1  & 46.2& 28.4 & 38.6 & 1.98& 0.65 & - & - & 15-30 & 21.6 \\
  \hline
 Cu  & 49.8  & 145.4& 11.94 & 41.5 & 2.92& 0.87 & - & - & 40-100& 54.7 \\
  \hline
Cu-Sn(3\%Sn)$^b$ & 56.1  & 135.2& 12.25 &42.3 & 2.41& 0.93 & - & - & 40-80$^d$ & 57.1 \\
\hline 
Cu-Sn(9\%Sn)$^c$ & 47.1  & 101.1& 12.94 & 39.6& 2.15& 0.72 & - & - & 40-80$^d$ & 42.9  \\
\hline
  Ni$_3$Al& 81.9& 179.8& 11.27&10.8 & 2.14&5.16& 1.88&1.613& 18.7-20.9~\cite{rigney1992}& 21.4 \\
 \hline
   FeAl& 95.0& 174.9& 11.78& 8.4&1.84& 4.98& 1.80&1.613& 16.6-25~\cite{specht1995}& 18.2 \\
    \hline 
 Ti$_3$Al& 62.6& 115.1& 16.51&7.2 &1.94&1.08& 1.38&1.613& 14-18& 11.1 \\
 \hline
   NiAl& 72.1& 162.5& 12.02& 4.5&2.25&1.43& 1.88&1.613& 6.4-7.1~\cite{ast2016}& 9.0 \\
 \hline
 TiAl& 74.8  & 109.8 & 16.15 &4.9 &1.46&2.26& 1.38&1.613& 8 & 8.6 \\
  \hline
  Al$_3$Sc& 65.8& 82.1& 17.32& 3.8& 1.24& 0.81& 1.613&1.19& 3.5~\cite{Gerberich1993}& 5.8 \\
\hline\hline
\end{tabular}
\end{table*}

By fitting the data of pure metals in Table II, we can get the form of the enhancement factor $\alpha$ as the function of $g(E_F)_R$:
\begin{equation}\label{E:enhancefactor}
 \alpha=43 \cdotp g(E_F)_R^{1/4} \,.
\end{equation}

Different from pure metals, intermetallics are composed of two or more elements with metallic bonding. Therefore, the interaction between different elements shall be considered in the model of fracture toughness. Electronegativity is a promising factor for this purpose as it describes the tendency of an atom to attract electrons thus forming localized states. For compound $A_m$$B_n$ we introduce an electronegativity factor as:
\begin{equation}\label{E:fen}
 f_{EN}=\beta/[1+ \frac{C^1_mC^1_n}{C^2_{m+n}}\sqrt{(\frac{(\chi_A-\chi_B)^2} {\chi_A \cdotp \chi_B} } ) ]^\gamma  \,,
\end{equation}
in which $C^1_m$, $C^1_n$ and $C^2_{m+n}$ refer to the number of combinations, $\chi_A$ and $\chi_B$ refer to the electronegativity of element A and B, respectively. The parameters $\beta$ and $\gamma$ can be obtained by fitting the data of intermetallics (See Table II) to be 0.3 and 8, respectively. In the above expression, both the degree of ionicity (the squared difference of electronegativity) and the strength of bonding (product of electronegativities) are taken into consideration. Therefore, we can get the enhancement factor $\alpha$ for pure metals and intermetallics as:
\begin{equation}\label{E:enhancefactor_total}
 \alpha=43 \cdotp g(E_F)_R^{1/4}\cdotp f_{EN} \,.
\end{equation}

We can plot the relation between the enhancement factor $\alpha$ to the relative DOS at the Fermi level $g(E_F)_R$ and electronegativity factor $f_{EN}$ (see Fig. 3). The enhancement factor $\alpha$ along the electronegativity factor axis decreases much faster than along the $g(E_F)_R$ axis. For instance, $g(E_F)_R$ and elastic properties of TiAl is comparable with pure metals, but its fracture toughness is much lower than pure metals. In this case, the electronegativity factor $f_{EN}$ plays an important role to determine the fracture toughness of such compounds.

\begin{figure}
  \centering
  \includegraphics[width=0.85\columnwidth]{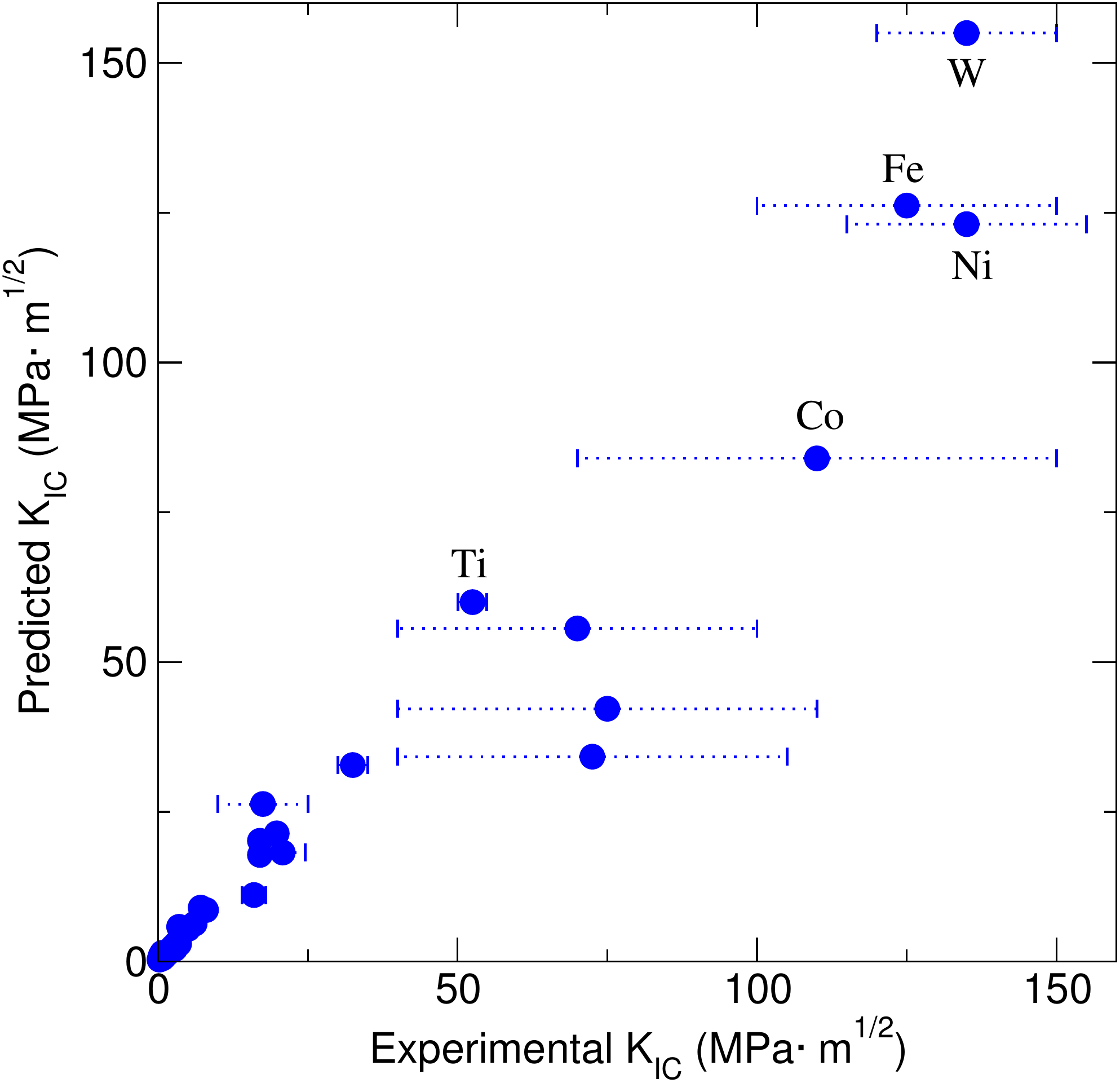}
  \caption{Comparison between experimental and predicted fracture toughness $K_{IC}$ for all the materials listed in Table I and Table II. The dotted lines along the $x$-axis refer to the experimental $K_{IC}$ distributions of metals.}
  \label{F:f4}
\end{figure}

The model of fracture toughness shown in Eq.~(\ref{E:KICmetal1}) is a universal model, which works for covalent and ionic crystals, metals and intermetallics. By using Eq.~(\ref{E:KICmetal1}) we have calculated fracture toughnesses for a series of metals and intermetallics as shown in Table II and plotted in Fig. 4. Experimental determination of fracture toughness can be affected by many factors. Taking into account the large spread of experimental values, our predicted results of $K_{IC}$ are in good agreement with experimental ones. Among all the metals in Table II, we can see that the predicted and experimental fracture toughnesses of Ni, Fe and W are higher than those of other metals, which explains why they are intrinsically suitable for mechanical applications. From Stone Age to Bronze Age to Iron Age, technological advances and human civilization were driven by the improvement of materials. From Table I and II we can see that the improvement of fracture toughness played a key role in the evolution of society.  

Finding materials with good comprehensive performance is always the key in materials design~\cite{ashby2004}. This is even more important when it comes to the mechanical properties that characterize strength (e.g., hardness) and wear-resistance (in particular, fracture toughness). With the establishment of the model of fracture toughness, we can guide the search of high performance materials through both theory and experiment. By evaluating the hardnesses of all the materials in Table I, we plot the hardness against fracture toughness of these materials (See Fig. 5). We can see that diamond, $c$-BN and WC are overall the best materials, which explains why they have played such an outstanding technological role. The remarkable hardness and counterintuitive high fracture toughness of diamond makes it irreplaceable in many areas, such as in mechanical processing area.

\begin{figure}
  \centering
  \includegraphics[width=0.9\columnwidth]{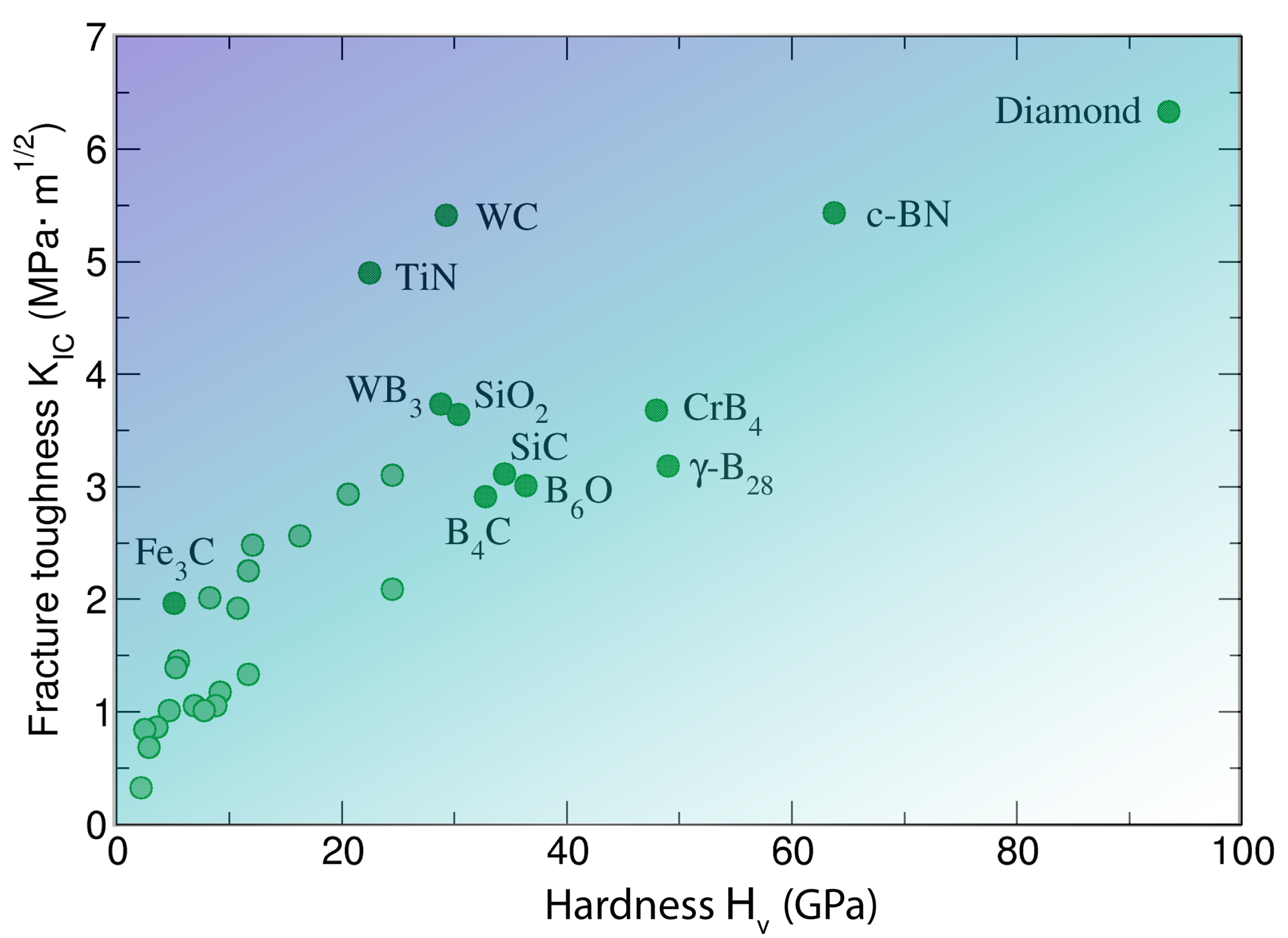}
  \caption{Correlation between hardness $H_v$ and fracture toughness $K_{IC}$.}
  \label{F:f5}
\end{figure}

In this letter, using the crystal structure information and properties derived from it, a simple and accurate fracture toughness model for covalent and ionic crystals has been constructed. Considering the intrinsic difference between covalent and ionic crystals and metals, we have introduced an enhancement factor $\alpha$, composed by the relative density of states at the Fermi level and atomic electronegativity of materials. The relative density of states at the Fermi level can measure the degree of metallicity, while the electronegativity factor takes into account the ionicity and strength of bonding. We have demonstrated that the model of fracture toughness is universal, which works for covalent and ionic crsytals, metals and intermetalllics. The predicted fracture toughness is in good agreement with the available experimental values. It is worth noting that all the parameters in the proposed fracture toughness model can be calculated directly and accurately by first-principles calculations, which makes the model applicable to a wide range of practical uses.

\begin{acknowledgments}
  We thank Alexander G. Kvashnin for useful discussions and Thomas Holvey for carefully reading the manuscript. The research was supported by the Russian Science Foundation (grant 17-73-20038).
\end{acknowledgments}



\begin{thebibliography}{99}

\bibitem{ashby2004} M. Ashby, \textit{Materials Selection in Mechanical Design} (Elsevier, Amsterdam, 2004).
\bibitem{ding2004} Z. Ding, S. Zhou, and Y. Zhao, Phys. Rev. B \textbf{70}, 184117 (2004).
\bibitem{king2007} S. King, and G. Antonelli, Thin Solid Films \textbf{515}, 7232 (2007).
\bibitem{ogata2009} S. Ogata and J. Li, J. Appl. Phys. \textbf{106}, 113534 (2009).
\bibitem{liu1989}  A. Y. Liu and M. L. Cohen, Science \textbf{245}, 841 (1989).
\bibitem{teter1998} D. M. Teter, MRS Bulletin \textbf{23}, 22 (1998).
\bibitem{niu2012} H. Niu, X.-Q. Chen, P. Liu, W. Xing, X. Cheng, D. Li, and Y. Li, Scientific Reports \textbf{2}, 718 (2012).
\bibitem{pugh1954} S. Pugh, Philos. Mag. \textbf{45}, 823 (1954).
\bibitem{chen2011} X.-Q. Chen, H. Niu, D. Li, and Y. Li, Intermetallics \textbf{19}, 1275 (2011).
\bibitem{zhang2013} X. Zhang, Y. Wang, J. Lv, C. Zhu, Q. Li, M. Zhang, Q. Li, and Y. Ma, J. Chem. Phys. \textbf{138}, 114101 (2013).
\bibitem{lyakhov2011} A. O. Lyakhov and A. R. Oganov, Phys. Rev. B \textbf{84}, 092103 (2011)
\bibitem{zhang2014-1} M. Zhang, M. Lu, Y. Du, L. Gao, C. Lu, and H. Liu, J. Chem. Phys. \textbf{140}, 174505 (2014).
\bibitem{szwacki2017} N. G. Szwacki, Scientific Reports \textbf{7}, 4082 (2017). 
\bibitem{gao2003} F. Gao, J. He, E. Wu, S. Liu, D. Yu, D. Li, S. Zhang, and Y. Tian, Phys. Rev. Lett. \textbf{91}, 015502 (2003). 
\bibitem{simunek2006} A. \v{S}im\u{u}nek, and J. Vack\'{a}\v{r}, Phys. Rev. Lett. \textbf{96}, 085501 (2006).
\bibitem{griffith1921} A. A. Griffith and M. Eng, Phil. Trans. R. Soc. Lond. A \textbf{221}, 163 (1921).
\bibitem{thomson1987} R. M. Thomson, J. Phys. Chem. Sol. \textbf{48}, 965 (1987).
\bibitem{nist1998} R. G. Munro, S. W. Freiman, and T. L. Baker, \textit{Fracture toughness data for brittle materials} (NIST, 1998)
\bibitem{perdew1996} J. P. Perdew, K. Burke, and M. Ernzerhof, Phys. Rev. Lett. \textbf{77,} 3865 (1996).
\bibitem{blochl1994} P. E. Blochl, Phys. Rev. B \textbf{50}, 17953 (1994).
\bibitem{kresse1993} G. Kresse and J. Hafner, Phys. Rev. B \textbf{47}, 558 (1993).
\bibitem{kresse1996} G. Kresse and J. Furthmuller, Phys. Rev. B \textbf{54}, 11169 (1996).
\bibitem{hill1952} R. Hill, Proc. Phys. Soc. \textbf{65}, 349 (1952).
\bibitem{cuadrado2011} N. Cuadrado, D. Casellas Padro, L. M. Llanes Pitarch, I. Gonzalez, and J. Caro, \textit{Proceedings of the Euro PM2011 Powder Metallurgy Congress and Exhibition} (2011).
\bibitem{cheng2014} X.-Y. Cheng, X.-Q. Chen, D.-Z. Li, and Y.-Y. Li, Acta Cryst. C \textbf{70}, 85 (2014).
\bibitem{weber2002} M. Weber, \textit{Handbook of Optical Materials, Laser and Optical Science and Technology} (Taylor and Francis, 2002)
\bibitem{anstis1981} G. Anstis, P. Chantikul, B. R. Lawn, and D. Marshall, J. Am. Ceram. Soc. \textbf{64}, 533 (1981).
\bibitem{tanaka1994} I. Tanaka, H.-J. Kleebe, M. K. Cinibulk, J. Bruley, D. R. Clarke, and M. Ruhle, J. Am. Ceram. Soc. \textbf{77}, 911 (1994).
\bibitem{lemaitre1988} P. Lemaitre, J. Mater. Sci. Lett. \textbf{7}, 895 (1988).
\bibitem{michot1988} G. Michot, A. George, A. Chabli-Brenac, and E. Molva, Scr. Metal. \textbf{22}, 1043 (1988).
\bibitem{tafreshi1995} M. J. Tafreshi, K. Balakrishnan, and R. Dhanasekaran, Mater. Res. Bulletin \textbf{30}, 1387 (1995).
\bibitem{ericson1988} F. Ericson, S. Johansson, and J.-A. Schweitz, Mater. Sci. Eng. A \textbf{105}, 131 (1988).
\bibitem{narita1987} N. Nrita, K. Higashida, and S. Kitano, Scr. Metal. \textbf{21}, 1273 (1987).
\bibitem{allen1989} L. C. Allen, J. Am. Chem. Soc. \textbf{111}, 9003 (1989).
\bibitem{ces2013} M. F. Ashby and D. Cebon, \textit{Cambridge Engineering Selector (CES)}  (Granta Design Limited, Cambridge, UK,
2013).
\bibitem{rigney1992} J. D. Rigney and J. J. Lewandowski, Mater. Sci. Eng. A \textbf{149}, 143 (1992).
\bibitem{specht1995} P. Specht and P. Neumann, Intermetallics \textbf{3}, 365 (1995).
\bibitem{ast2016} J. Ast, B. Merle, K. Durst, and M. Goken, J. Mater. Res. \textbf{31}, 3786 (2016).
\bibitem{Gerberich1993} W. Gerberich, S. Venkataraman, J. Hoehn, and P. Marsh, \textit{Structural intermetallics} (Pennsylvania, 1993).


\end{thebibliography}
\end{document}